\documentclass[
    ,final            
  ]
  {aipproc}

\layoutstyle{8x11double}

\renewcommand\XFMtitleblock{%
  \XFMtitle
  \let\XFMoldpar\par
  \def\par{\XFMoldpar\def\par{\space 
             (on behalf of the MAGIC Collaboration)\XFMoldpar}}%
   \XFMauthors
   \let\par\XFMoldpar
   \XFMaddresses
   \XFMabstract
   \vspace{5pt}%
   \XFMkeywords
   \XFMclassification
 }

\begin{document}
 
\title{Results of MAGIC on Galactic sources}


\author{Javier Rico}{
  address={Instituci\'o Catalana de Recerca i Estudis Avancats (ICREA) \& \\
Institut de F\'{\i}sica d'Altes Energies (IFAE) \\
Edifici Cn. Universitat Aut\`onoma de Barcelona \\
08193 Bellaterra (Barcelona) Spain }
}

\begin{abstract}
MAGIC is a single-dish Cherenkov telescope located on La Palma
(Spain), hence with an optimal view on the Northern sky. Sensitive in
the 30 GeV -- 30 TeV energy band, it is nowadays the only ground-based
instrument being able to measure high-energy $\gamma$-rays below 100
GeV. We review the most recent experimental results on Galactic
sources obtained using MAGIC. These include pulsars, binary systems,
supernova remnants and unidentified sources.
\end{abstract}

\maketitle


\section{Introduction: the MAGIC telescope} 

The Major Atmospheric Gamma Imaging Cherenkov (MAGIC) telescope is a
last-generation instrument for very high energy (VHE, $E \geq
50-100$~GeV) $\gamma$-ray observation exploiting the Imaging Air
Cherenkov (IAC) technique. It is located on the Roque de los Muchachos
Observatory ($28^\circ 45^\prime 30^{\prime\prime}$N, $17^\circ$
$52^\prime$ $48^{\prime\prime}$W, 2250~m above see level) in La Palma
(Spain). This kind of instrument images the Cherenkov light produced
in the particle cascade initiated by a $\gamma$-ray in the
atmosphere. MAGIC incorporates a number of technological improvements
in its design and is currently the largest single-dish telescope
(diameter 17~m) in this energy band, yielding the lowest threshold
(55~GeV with the nominal trigger, 25~GeV with the pulsar trigger).
Since February 2007, MAGIC signal digitization has been upgraded to
2\,GSample/s Flash Analog-to-Digital Converters (FADCs), and timing
parameters are used during the data analysis~\cite{Tescaro2007}. This
results in an improvement of the flux sensitivity from 2.5$\%$ to
1.6$\%$ (at a flux peak energy of 280\,GeV) of the Crab Nebula flux in
50 hours of observations. The relative energy resolution above 200~GeV
is better than 30$\%$. The angular resolution is $\sim 0.1^\circ$,
while source localization in the sky is provided with a precision of
$\sim 2'$. MAGIC is also unique among IAC telescopes by its capability
to operate under moderate illumination~\cite{moon} (i.e.\ moonlight
and twilight). This allows to increase the duty cycle by a factor 1.5
and a better sampling of variable sources is possible. The
construction of a second telescope is now in its final stage and MAGIC
will start stereoscopic observations in the near future.

\section{Recent Results on Galactic Objects} 

The physics program developed with the MAGIC telescope includes both,
topics of fundamental physics and astrophysics. Regarding Galactic
observations, MAGIC has discovered four new VHE $\gamma$-ray
sources~\cite{lsi,ic443,cygX1,crabpulsar} and studied in detail eight
of the previously
known~\cite{wr147,tev,casA,crab,psr1951,sgrA,hess1834,hess1813}. In
this paper we highlight our latest contributions to Galactic
astrophysics. The results from extragalactic observations are
presented elsewhere in these proceedings~\cite{wagner}.

\subsubsection{TeV 2032+4130}

\begin{figure}[!t]
\centering
\includegraphics*[angle=0,width=0.8\columnwidth]{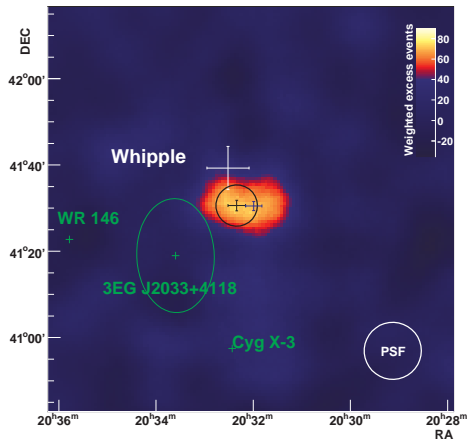}
\caption{Skymap of $\gamma$-ray candidate events
(background-subtracted) for energies above 500~GeV. The MAGIC position
is shown with a black cross. Also shown are the last positions reported
by Whipple and HEGRA.}
\label{fig:tev} 
\end{figure}

\begin{figure}[!t]
\centering
\includegraphics*[angle=0,width=\columnwidth]{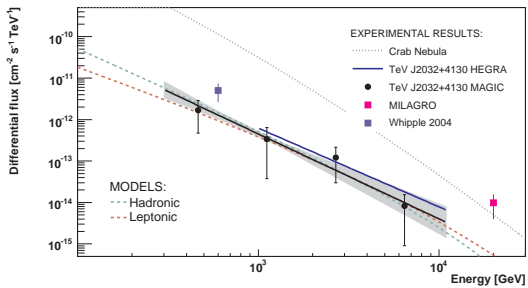}
\caption{
Differential energy spectrum from TeV J2032+4130. The shaded area
shows the 1$\sigma$ error in the fitted energy spectrum. The flux
observed by Whipple in 2005 and in the Milagro scan are marked with
squares. The light line shows the HEGRA energy spectrum. Theoretical
one-zone model predictions are depicted with dashed lines.}
\label{fig:tevspec} 
\end{figure}

The TeV source J2032+4130 was the first unidentified very high energy
(VHE) $\gamma$-ray source, and also the first discovered extended TeV
source, likely to be Galactic~\cite{tevhegra}. The field of view of
TeV~J2032+4130 was observed with MAGIC for 93.7 hours of good-quality
data, between 2005 and 2007~\cite{tev}. The source is extended with
respect to the MAGIC PSF (see Figure~\ref{fig:tev}). Its intrinsic
size assuming a Gaussian profile is $\sigma_{\rm src}$
=5.0$\pm$1.7$_{\rm sta}\pm$0.6$_{\rm sys}$ arcmin. The energy spectrum
is well fitted ($\chi^2/n.d.f=0.3$) by the following power law:
$\frac{dN}{dE dA dt} = (4.5\pm
0.3)\times10^{-13}(E/1~TeV)^{-2.0\pm0.3}$
TeV$^{-1}$cm$^{-2}$s$^{-1}$. Quoted errors are statistical, the
systematic error is estimated to be 35$\%$ in the flux level and 0.2
in the photon index~\cite{crab}. The MAGIC energy spectrum (see
Figure~\ref{fig:tevspec}) is compatible both in flux level and photon
index with the one measured by HEGRA, and extends it down to
400~GeV. We do not find any spectral break, nor any flux variability
over the 3 years of MAGIC observations.

\subsubsection{Wolf-Rayet binaries}

WR stars display some of the strongest sustained winds among galactic
objects with terminal velocities reaching up to $v_\infty
>1000-5000$\,km/s and also one of the highest known mass loss rate
$\dot M \sim 10^{-4}...10^{-5}~M_\odot/$\,yr.  Colliding winds of
binary systems containing a WR star are considered as potential sites
of non-thermal high-energy photon production, via leptonic and/or
hadronic process after acceleration of primary particles in the
collision shock (see, e.g., \cite{Reimer06}).

\begin{figure}[t!]
\centering
\includegraphics*[angle=0,width=6.5cm]{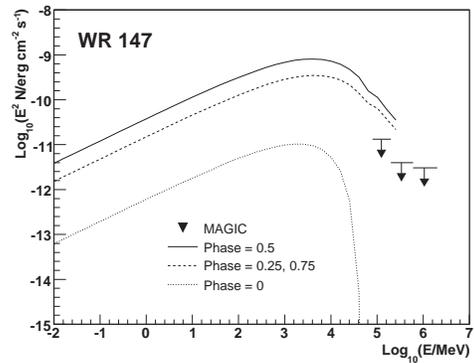}
\caption{Inverse Compton (IC) spectra of WR\,147 for orbital phases 0, 0.25,
0.5 and 0.75~\cite{Reimer06} together with MAGIC experimental upper
limits.}
\label{fig:wr}
\end{figure}

We have selected two objects of this kind, namely WR\,147 and WR\,146,
and observed them for 30.3 and 44.5 effective hours,
respectively~\cite{wr147}. No evidence for VHE $\gamma$-ray emission has
been detected in either case, and upper limits to the emission of 1.5,
1.4 and 1.7$\%$ (WR\,147) and 5.0, 3.5 and 1.2$\%$ (WR\,146) of the
Crab Nebula flux are derived for lower energy cuts of 80, 200 and 600
GeV, respectively. These limits are shown in Figure~\ref{fig:wr} for
the case of WR\,147, compared with a theoretical
model~\cite{Reimer06}.

\subsubsection{Cassiopeia A}

\begin{figure}[!tp]
\centering
\includegraphics*[angle=0,width=\columnwidth]{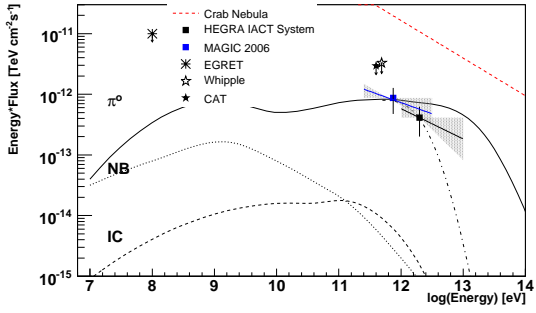}
\caption{\label{fig:casA} Spectrum of Cas~A as measured by MAGIC. The
upper limits given by Whipple, EGRET and CAT are also indicated, as
well as the HEGRA detection. The MAGIC and HEGRA spectra are shown in
the context of the model by \cite{Berezhko}.}
\end{figure}

We observed the shell-type supernova remnant (SNR) Cassiopeia\,A
during 47 good-quality hours, and detected a point-like source of VHE
$\gamma$-rays above $\sim$250~GeV~\cite{casA}. The measured spectrum
is consistent with a power law with a differential flux at 1~TeV of
(1.0$\pm$0.1$_{stat}\pm$0.3$_{sys})\times10^{-12}$
TeV$^{-1}$cm$^{-1}$s$^{-1}$ and a photon index of
$\Gamma$=2.4$\pm$0.2$_{stat}$$\pm$0.2$_{sys}$. The spectrum measured
about 8 years later by MAGIC is consistent with that measured by
HEGRA~\cite{casAhegra} for the energies above 1~TeV, i.e, where they
overlap (see Figure~\ref{fig:casA}). Our results seem to favor a
hadronic scenario for the $\gamma$-ray production, since a leptonic
origin of the TeV emission would require low magnetic field
intensities, which is in principle difficult to reconcile with the
high values required to explain the rest of the broad-band
spectrum. However, hadronic models~\cite{Berezhko} predict for the 100
GeV -- 1 TeV region a harder spectrum than then measured one.

\subsubsection{IC 443/MAGIC~J0616+225}

\begin{figure}[!t]
\centering
\includegraphics*[angle=0,width=0.9\columnwidth]{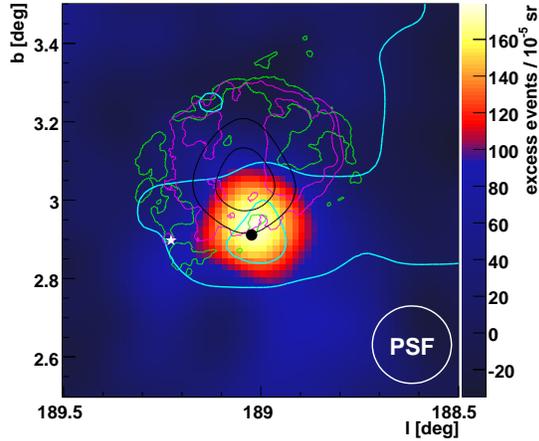}
\caption{\label{fig:ic443} 
Sky map of $\gamma$-ray candidate events (background subtracted) in the
direction of MAGIC J0616+225 for an energy threshold of about 150
GeV. Overlayed are $^{12}$CO emission contours, contours of 20 cm VLA
radio data, X-ray contours and $\gamma$-ray contours from EGRET.  The
white star denotes the position of the pulsar CXOU
J061705.3+222127. The black dot shows the position of the 1720 MHz OH
maser.}
\end{figure}

\begin{figure}[!ht]
\includegraphics*[angle=0,width=0.7\columnwidth]{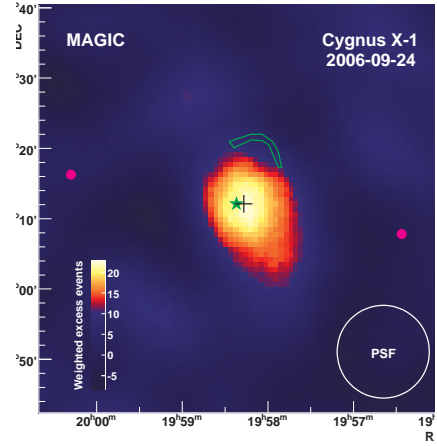}
\caption{Skymap of $\gamma$-ray excess events (background subtracted)
above 150 GeV around Cygnus X-1 corresponding to the flare detected on
2006-09-24. The cross shows the best-fit position of the $\gamma$-ray
source. The position of the X-ray source and radio emitting ring-like
are marked by the star and contour, respectively.\label{fig:cygX1}}
\end{figure}

We have detected a new source of VHE $\gamma$-rays located close to
the Galactic Plane, namely MAGIC\,J0616+225~\cite{ic443}, which is
spatially coincident with the SNR IC\,443. The measured energy
spectrum is well fitted ($\chi^2/n.d.f=1.1$) by the following power
law: $\frac{dN}{dE dA dt} = (1.0\pm
0.2)\times10^{-11}(E/0.4~TeV)^{-3.1\pm0.3}$
TeV$^{-1}$cm$^{-2}$s$^{-1}$. MAGIC\,J0616+225 is point-like for MAGIC
spatial resolution, and appears displaced to the south of the center
of the SNR shell, and correlated with a molecular cloud~\cite{cornett}
and the location of maser emission~\cite{claussen} (see
Figure~\ref{fig:ic443}). There is also an EGRET source centered in the
shell of the supernova remnant. The observed VHE radiation may be due
to $\pi^0$-decays from interactions between cosmic rays accelerated in
IC\,443 and the dense molecular cloud. A possible distance of this
cloud from IC\,443 could explain the steepness of the measured VHE
$\gamma$-ray spectrum.

\subsubsection{Cygnus X-1}

Cygnus X-1 is the best established candidate for a stellar mass
black-hole (BH) and one of the brightest X-ray sources in the sky. We
have observed it for 40 hours along 26 different nights between June
and November 2006. Our observations have imposed the first limits to
the steady $\gamma$-ray emission from this object, at the level of 1$\%$
of the Crab Nebula flux above $\sim$500\,GeV. We have also obtained a
very strong evidence (4.1$\sigma$ post-trial significance) of a
short-lived, intense flaring episode during 24th September 2006, in
coincidence with a historically high flux observed in
X-rays~\cite{malzac} and during the maximum of the $\sim 326$d
super-orbital modulation~\cite{rico}. The detected signal is
point-like, consistent with the position of Cygnus X-1. The nearby 
radio-nebula produced by the jet interaction with the interstellar
medium~\cite{Gallo} is excluded as the possible origin of an eventual
putative emission (see Figure~\ref{fig:cygX1}).

\subsubsection{LS I +61 303}

\begin{figure}[!t]
\includegraphics*[angle=0,width=7cm]{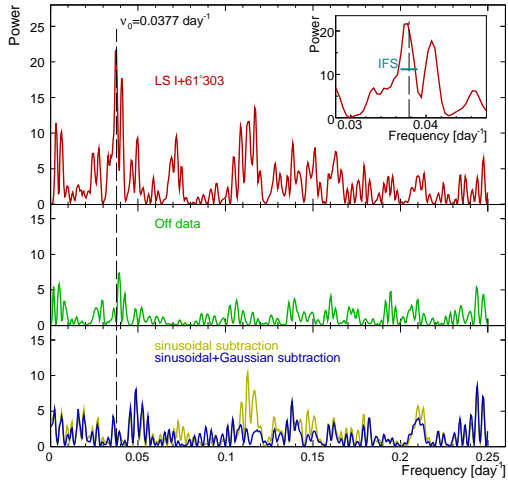}
\caption{Lomb-Scargle periodogram for LS\,I\,+61\,303 data (upper
panel) and simultaneous background data (middle panel). In the lower
panel we show the periodograms after subtraction of a sinusoidal
signal at the orbital period (light line) and a sinusoidal plus a
Gaussian wave form (dark line). The vertical dashed line corresponds
to the orbital frequency. Inset: zoom around the highest peak, which
corresponds to the orbital frequency (0.0377d$^{-1}$). Its post-trial
probability is $\sim$10$^{-7}$.  The IFS is also shown.
\label{fig:lsi}}
\end{figure}

LS\,I\,+61\,303 is a very peculiar binary system containing a
main-sequence star together with a compact object (neutron star or
black hole), which displays periodic emission throughout the spectrum
from radio to X-ray wavelengths. Observations with MAGIC have
determined that this object produces $\gamma$-rays up to at least $\sim$4
TeV~\cite{lsi}, and that the emission is periodically modulated by the
orbital motion ($P_\textrm{TeV}=(26.8 \pm 0.2)$\,d)~\cite{lsiperiodic}
(see Figure~\ref{fig:lsi}). The peak of the emission is found always
at orbital phases around 0.6--0.7. During December 2006 we detected a
secondary peak at phase 0.8--0.9. Between October-November 2006, we
set up a multiwavelenght campaign involving radio (VLBA, e-EVN,
MERLIN), X-ray (Chandra) and TeV (MAGIC) observations~\cite{lsimw}.
We have excluded the existence of large scale ($\sim 100$
mas) persistent radio-jets, found a possible hint of
temporal correlation between the X-ray and TeV emissions and evidence
for radio/TeV non-correlation.

\subsubsection{Crab Nebula}

\begin{figure}[!t]
\includegraphics*[angle=0,width=7cm]{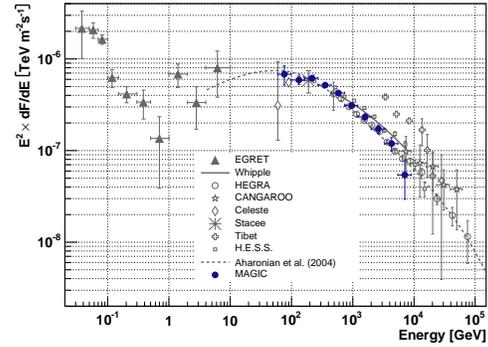}
\caption{
SED of the $\gamma$-ray emission of Crab Nebula. The measurements shown
below 10GeV are by EGRET. In VHE $\gamma$-rays, measurements are from
ground-based experiments. 
\label{fig:crab}}
\end{figure}

The Crab Nebula is the standard candle for VHE astrophysics and as
such, a big fraction of MAGIC observation time is devoted to this
object. Out of it, we have used 16 hours of optimal data to measure
the energy spectrum between 60 GeV and 8 TeV~\cite{crab}. The peak of
the SED has been measured at an energy $E=(77\pm 35)$\,GeV (see
Figure~\ref{fig:crab}). The VHE source is point-like and the position
coincides with that of the pulsar. More recently, thanks to a special
trigger setup, we have detected pulsed emission coming from the Crab
pulsar above 25\,GeV, with a statistical significane of
6.4$\sigma$~\cite{crabpulsar}. This is the first time that a pulsed
$\gamma$-ray emission is detected from a ground-based telescope, and
opens the possibility of a detailed study of the pulsar's energy
cutoff, which will help elucidate the mechanism of high energy
radiation in these objects.

\vspace{0.5cm}

We thank the Instituto de Astrofisica de Canarias for the excellent
working conditions at the Observatorio del Roque de los Muchachos in
La Palma. 

\end{document}